\documentclass[aps,prd]{revtex4}
\usepackage{graphicx}
\usepackage{color}
\usepackage{slashed}
\usepackage{amsmath}

\newcommand\bef{\begin{figure}}
\newcommand\eef[1]{\label{fg:#1}\end{figure}}
\newcommand\beq{\begin{equation}}
\newcommand\eeq[1]{\label{#1}\end{equation}}
\newcommand\beqa{\begin{eqnarray}}
\newcommand\eeqa[1]{\label{#1}\end{eqnarray}}
\newcommand\bet{\begin{table}}
\newcommand\eet[1]{\label{tb:#1}\end{table}}

\newcommand\fgn[1]{Figure \ref{fg:#1}}
\newcommand\eqn[1]{eq.\ (\ref{#1})}

\newcommand\tbn[1]{Table \ref{tb:#1}}

\newcommand\ie{{\sl i.e.\/}}

\newcommand\etal{{\sl et al.\/}}

\newcommand{\cfr}{{\rm CFR\/}}
\newcommand{\ifr}{{\rm IFR\/}}

\begin{document}

\title{Estimating the number of COVID-19 infections in Indian hot-spots using fatality data}

\author{Sourendu Gupta}
\affiliation{Department of Theoretical Physics, Tata Institute of Fundamental
 Research, Homi Bhabha Road, Mumbai 400005, India}
\author{R.\ Shankar}
\affiliation{The Institute of Mathematical Sciences, CIT campus, Chennai 60013,
India}
\begin{abstract}
In India the COVID-19 infected population has not yet been accurately
established. As always in the early stages of any epidemic, the need to
test serious cases first has meant that the population with asymptomatic
or mild sub-clinical symptoms has not yet been analyzed.  Using counts
of fatalities, and previously estimated parameters for the progress of
the disease, we give statistical estimates of the infected population.
The doubling time, $\tau$, is a crucial unknown input parameter which
affects these estimates, and may differ strongly from one geographical
location to another.  We suggest a method for estimating epidemiological
parameters for COVID-19 in different locations within a few days, so
adding to the information required for gauging the success of public
health interventions. \\
TIFR/TH/20-10 (DRAFT)
\end{abstract}
\maketitle

\section{Introduction}

It is generally accepted that in many parts of the world the actual number
of infected people is much more than the number of confirmed cases. This
is due to limited testing which biased towards the serious cases. The
number of documented fatalities on the other hand is likely to be much
closer to the actual number. In this note we use a method to estimate
the actual number of infections from the documented number of fatalities.
This estimate is one of our main results. It is important because it is
seems possible that asymptomatic and sub-clinical infections may also
infect others \cite{news}. We find large uncertainties in these predictions
at this time, and suggest a method to improve the estimates systematically
day by day. This gives us a secondary motivation, which is to refine
epidemiological parameters for COVID-19 infections using only the daily
statistics of fatalities.

We do not utilize detailed epidemiological models. Our model input is a
hypothesis of exponential growth. There is no data from India at present
which contradicts this. The other inputs we need are about the progress
of the disease. There is agreement in the literature that post-infection
there is a short asymptomatic period. We use statistical models
for the progression of the disease from asymptomatic to resolution into
recovery or fatality which are parametrized to fit reports. We also need
the infection fatality ratio (IFR), which we take from previous studies.
Using these we make predictions for the infected population now and in
future for various scenarios for the exponential growth rate. We discuss
how our predictions can be used to validate some of the model assumptions.

The distribution of population in India is highly non-uniform, and this
could cause geographical fluctuations in the progress of the epidemic.
So we apply our estimators to various localized outbreaks seen in
India till now, and make predictions for the number of fatalities in
these regions for about a week from now.  Using the statistical inputs
which we discuss here, we give confidence limits on the predictions,
and discuss how to match them to future data in order to extract the
remaining epidemiological parameters. It should be noted that current
data is available aggregated over districts. Finer details may be very
useful for discussing exit strategies from a lock-down. In view of this, the
availability of data from each hospital separately would be of great use.

\section{Inputs}

\subsection{Model for growth of infections}

In the early stages of a typical epidemic, when the number infected are a very
small fraction of the population, the number of infected cases, $I$, rises
exponentially. This may be parametrized as
\beq
  I(t) = I(t_0) {\rm e}^{\lambda (t-t_0)} = I(t_0) 2^{(t-t_0)/\tau},
\eeq{exp}
where the doubling time $\tau=\ln2/\lambda$, and $t_0$ is the
initial time at which the counting starts.

The doubling time $\tau$ is related to the basic reproductive rate
parameter $R_0$. This is affected by both the virulence of the pathogen
and the rate of social contact. Estimates of $R_0$ in China vary from
as low as 2.2 to as high as 14.8, with a cluster of estimates close to
the median of 2.79 \cite{repraterev}. See also an interesting estimate
of the effect of public health interventions on $R_0$ using data from
the cruise ship Diamond Princess \cite{diamond}. Converting $R_0$ to
$\tau$ requires an epidemiological model. In this note, we do not use
any particular model. Our only assumption about the dynamics of the
epidemic is the exponential growth of infections given by \eqn{exp}.

Due to the extreme heterogeneity of the population in India, $R_0$,
or $\tau$, could vary from one place to another. The most definitive
estimates of population densities are almost a decade old, since they come
from the 2011 census of India \cite{census}. According to this source,
the population density of Indore is 9718 $/{\rm Km\/}^2$, whereas that
for Mumbai is 28,185 $/{\rm Km\/}^2$. Inside Mumbai again there is
extreme heterogeneity, with high density areas like Dharavi having a
density of 3,35,743 $/{\rm Km\/}^2$. These numbers are indicative of
the possibility that different cities, and even districts within a city,
may have very different doubling times $\tau$.

In view of this, we do not assume a single value for $\tau$, but work
with two scenarios: in Scenario A infections double every 10 days, \ie,
$\tau=10$ days; in Scenario B doubling is twice as rapid, with $\tau=5$
days. In comparison, the aggregated data on fatalities in India taken
until 30 March, indicates a doubling time of roughly 4 days. We note
that between March 30 and April 2, the number of fatalities in the
Mumbai-Navi Mumbai-Pune area changed from 8 to 14, which supports the
idea that Scenario B may not be far from the current doubling time in
this location. The aggregated Indian fatality numbers could either be
dominated by rapid growth in some local outbreaks, or a general spread
of the disease. The geographically disaggregated data is capable of
indicating which is the case.

\subsection{Model for disease progression}

For COVID-19 infections the incubation period, $t_i$, is estimated
to be 5.1 days (95\% CL $4.5$--$5.8$ days), with a long tail
\cite{incubation}. Early observers reported that the recovery time, $t_r$,
the time from the onset of symptoms to recovery, ranged from 12 to 32
days \cite{crt}. The interval from onset of symptoms to release from
hospital for a recovery may depend on various factors.  It is seen to
be larger from the time to death. This latter number is seen to have a
mean of 17.8 days \cite{ifr}, and is the one relevant for our analysis.
We will take the duration of the disease, $T$, to be the sum of these
two periods, \ie, $T=t_i+t_r$. We will take the incubation period to be
a minimum of 3 days and have an exponential distribution after that, so
that the mean incubation period is 5.1 days. We shall take the recovery
time to have a minimum of 12 days, and be exponentially distributed
such that the mean time to death is 17.8 days. In other words, $t_i$
and $t_r$ are random variates chosen from the distribution
\beq
  P(t_i)= \begin{cases}
     0 & {\rm for\ } t_i<3,\\
     2.1\,{\rm e}^{-(t_i-3)/2.1} & {\rm for\ } t_i>3\\
    \end{cases}
  \qquad{\rm and}\qquad
  P(t_r)= \begin{cases}
     0 & {\rm for\ } t_r<12,\\
     5.8\,{\rm e}^{-(t_r-12)/5.8} & {\rm for\ } t_i>12.\\
    \end{cases}
\eeq{ptiptr}
With these assumptions, the mean infected period, $T$, is about 19 days,
with the 95\% CL being 15--27 days. These distributions can be improved
in future. The core point to note is that a Gaussian distribution
is not a good description when the data shows a long tail and skewness.
Note that each of the random variates is applicable to a different case.

\subsection{Model for the chance of fatality}

The fraction of the infected population which dies is called the
infection fatality ratio, $\ifr$. This is most reliably estimated after
the end of an epidemic. Estimates based on Chinese data for COVID-19
give $\ifr=0.0066$ (\ie, $0.66$\%) on average \cite{ifr}, with a strong
age structure \cite{ifr,cfr,agestr}. The analysis of \cite{ifr} gives a
skewed posterior distribution of this quantity, so We will take $\ifr$
to have a minimum value of 0.0035, and exponentially after this so that
it has a mean of $0.0066$ and a width of $0.005$.  
\beq
  P(\ifr)= \begin{cases}
     0 & {\rm for\ } \ifr<0.0035,\\
     0.00307\,{\rm e}^{-(\ifr-0.0035)/0.00307} & {\rm for\ } \ifr>0.0035\\
    \end{cases}
\eeq{ifr}
Since the case fatality ratio, $\cfr$, is discussed more frequently, we
discuss the relation between $\cfr$ and $\ifr$ in the final section.

Note that $\ifr$ is a population averaged quantity, and the random values
assumed for it are a measure of our uncertainty about this number. We have
assumed that there is no correlation between $T$ and $\ifr$, since we have
not taken age structuring into account. In a more detailed, age structured
analysis, correlations may be important. 

\subsection{Data for India}

The remainder of our analysis will use the data on fatalities reported by
the Ministry of Health and Family Welfare \cite{mohfw}. Before embarking
on the analysis it is useful to separate out the data on fatalities into
two sets. One is the set of fatalities known to be of persons who arrived
from a foreign country and was very soon after diagnosed as being infected
with COVID-19. The statistics of such deaths relate to infections in the
country where it was picked up. So this set is not of relevance to our
analysis of the infections within the country. It is the complement, namely
the set of fatalities to which no travel history can be attributed,
which is of relevance to the analysis. This separation is not made
in \cite{mohfw}.  However, the data tracked in \cite{wiki} adds this
information from press reports, and the totals tally with the data of
\cite{mohfw}. This is the data set we utilize here\footnote{We make our
data set on clusters of fatalities publicly available on Google maps at
the address drive.google.com/open?id=10lVH6aWD1ulwvLnaC-MCDwD1jnsci6H6}.

Due to the extreme variability in population density across India, it
is good to avoid a country-averaged analysis if possible.  We analyze
clusters of fatalities due to COVID-19 in India, with data that was
complete at the end of March 30, 2020.  The fatalities of persons with no
history of recent foreign travel fell into two groups: sporadic, defined
as single fatalities in isolated geographical locations, and clusters,
defined as more than one fatality in the same city or in towns very close
to each other. We decided not to use sporadic cases, since statistical
estimates are meaningless for single incidents. We found four clusters
which are listed in \tbn{stats}. These are the epidemic hot-spots in the
country according to currently available data.

\section{Methods and Results}

The observation of the number of fatalities, $D(t)$, on day $t$, may be
converted to an estimate of the actual number of infections, $I(t)$, on
the same day by using the formula
\beq
  I(t) = \left[\frac{D(t)}{\ifr}\right] 2^{(t-T)/\tau}
\eeq{estimator}
The first factor, $D(t)/\ifr$ is an estimate of the number of infections,
$I(t-T)$, on day $t-T$. The second factor is the exponential growth of
\eqn{exp} which evolves this older number of infections to its current
value, leading to \eqn{estimator}. 

\bef
\begin{center}
 \includegraphics[scale=0.6]{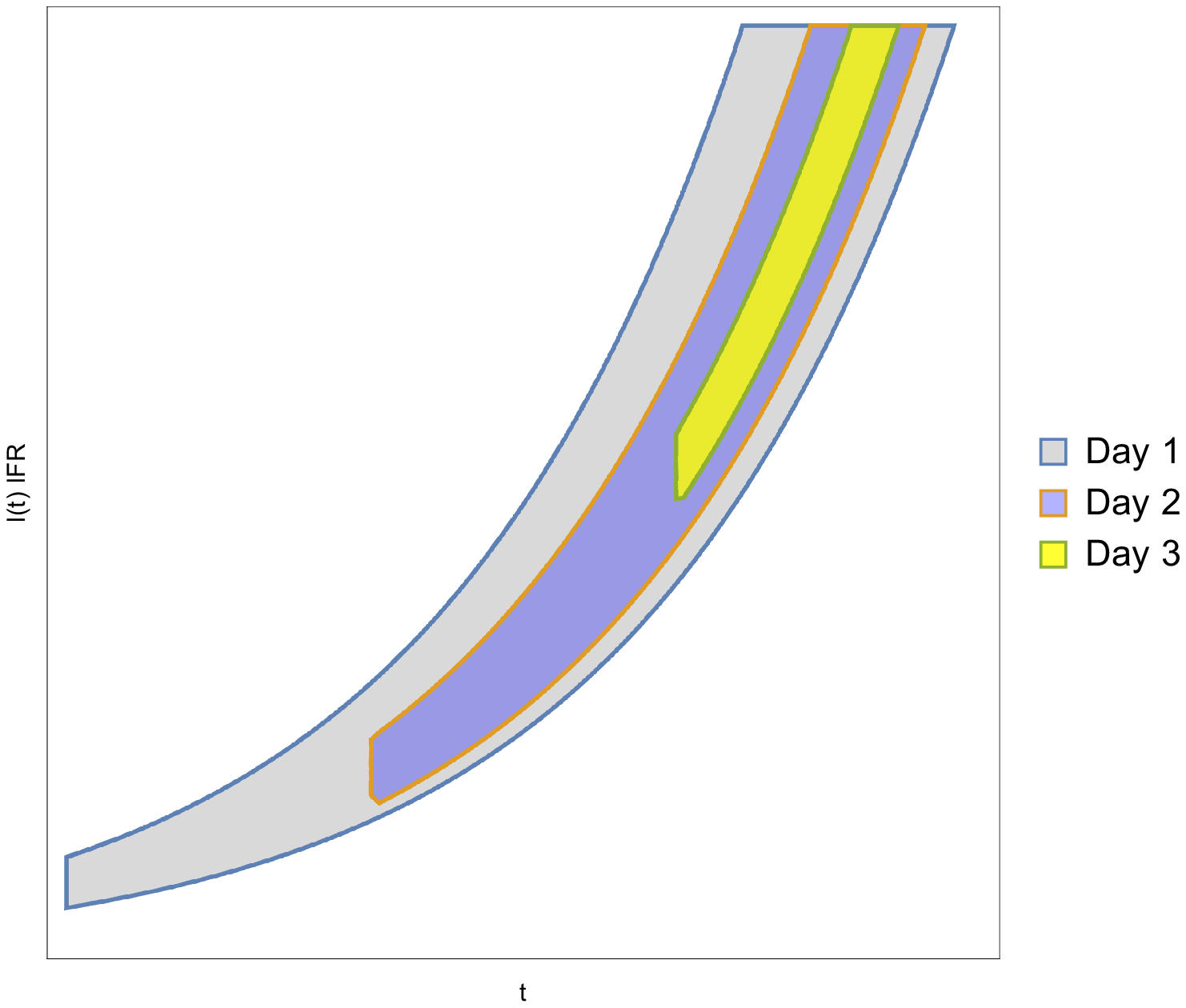}
\end{center}
\caption{A scheme for the successive improvement of the predicted
  ${\cal D}(t)$ with successive days' data on $D(t)$. Each day, the
  observed value of $D(t)$ is extrapolated back to the initial time
  to narrow the initial predicted interval of $I(t)$ through a
  Bayesian procedure.}
\eef{bayes}

Note that the very broad and skewed distributions of $T$ and $\ifr$ will
give similarly broad and skewed distributions of $I(t)$. Here we suggest
how to narrow these estimates progressively. Any knowledge of $I(t)$
gives a prediction of $I(t')$ at a future date $t'$. Uncertainties in
$I(t)$ expand into larger uncertainties in $I(t')$ due to exponential
growth. However, the number of fatalities up to date $t'$, \ie. $D(t')$,
is directly observable. A time series for $D(t')$ allows us to estimate
$\tau$ directly. Furthermore, with each day's data on fatalities, one can
run the evolution backwards to rule out some of the uncertainty in the
starting prediction $I(t)$ on 30 March. This means that the prediction for
${\cal D}(t')=I(t')\times\ifr$ further in the future is narrowed down.
We show this inference procedure schematically in \fgn{bayes}.  As the
allowed range of the initial $I(t)$ successively narrows, one also narrows
the allowed range of $T$ and $\ifr$ through a Bayesian inference paradigm.

\bef
\begin{center}
 \includegraphics[scale=0.6]{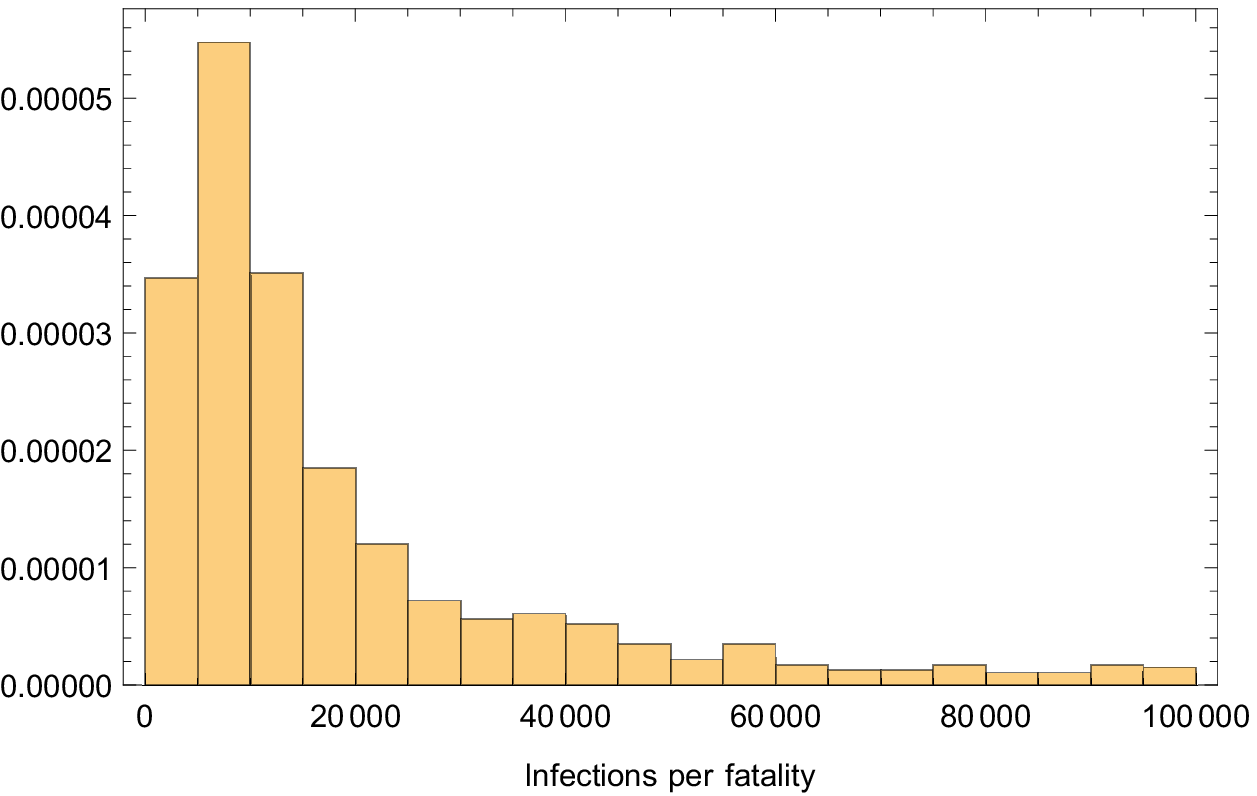}
 \includegraphics[scale=0.6]{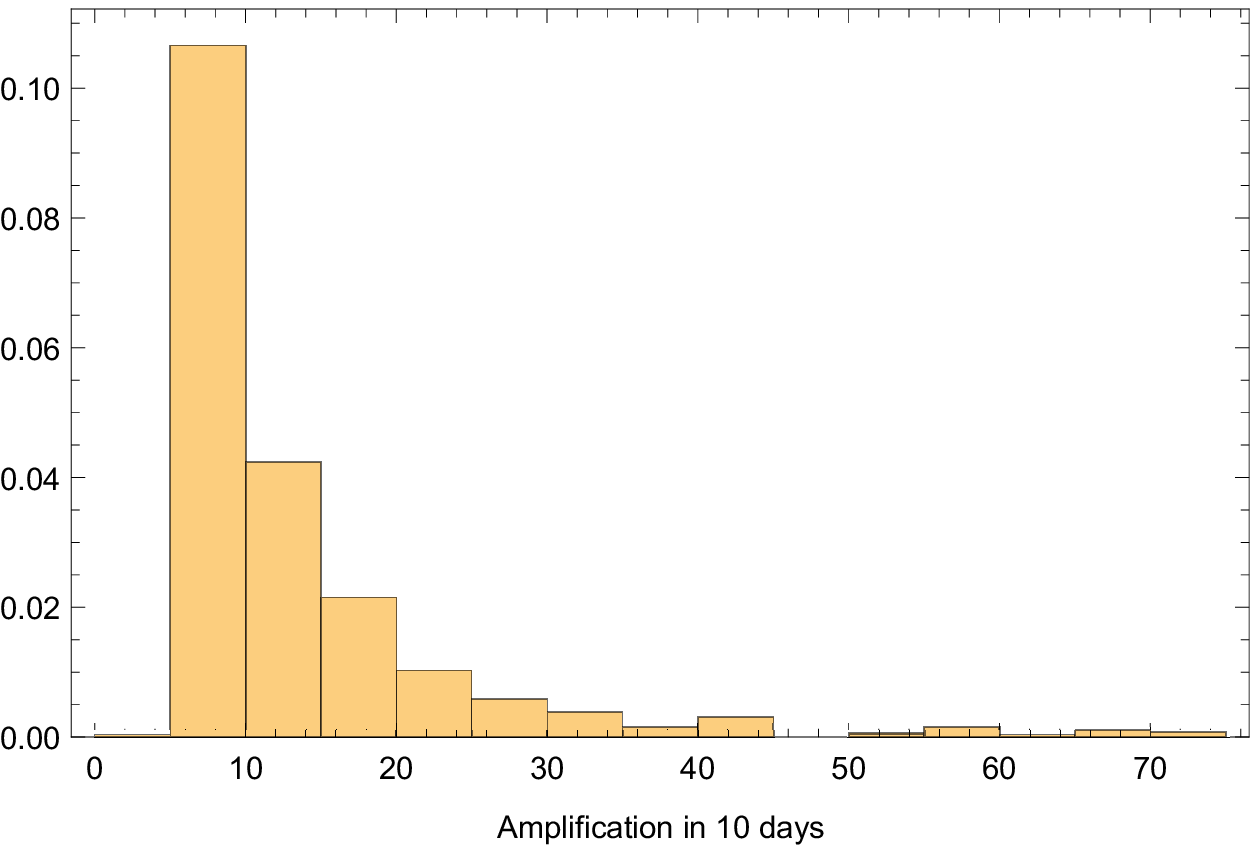}
\end{center}
\caption{We show the probability distraction of the ratio $I(t)/D(t)$
 (histogram on the left) and ${\cal D}(t+10)/D(10)$ (histogram on the right) in
 scenario B. The 95\% confidence limit from the distribution on the left
 is the estimate quoted in \eqn{estimate}.}
\eef{distrib}

There are statistical uncertainties in the parameters $T$ and $\cfr$.
We have combined them through a random Monte Carlo sampling using
the probability distribution functions defined above. We implemented
the Monte Carlo in Mathematica. The use of
\eqn{estimate} to make estimates of $I(t)$, and the future values $I(t')$
and $D(t')$, then gives a statistical distribution of these quantities.
These are implemented in the same Monte Carlo estimator. The basic result
is for the amplification factor, $I(t)/D(t)$, which we obtain with this
numerical estimate,
\beq
  I(t) = 
   \begin{cases}
    \left(460\;{\rm to}\;9,700\right)\times D(t) & {\rm for\ scenario\ A}\\
    \left(2,700\;{\rm to}\;3,80,000\right)\times D(t) & {\rm for\ scenario\ B}\\
   \end{cases}
\eeq{estimate}
The ranges given here are 95\% CL, and have been rounded to two significant
digits. The median values are 1,500 in scenario A and 12,000 in scenario B.

We tested the effect of changing the skewness of the distribution by replacing
the shifted Exponential distributions in \eqn{ptiptr} and \eqn{ifr} by Gamma
distributions tuned so as to reproduce the same means and variances. However,
the Gamma distributions then have smaller skewness.
\beq
  I(t) = 
   \begin{cases}
    \left(440\;{\rm to}\;7,700\right)\times D(t) & {\rm for\ scenario\ A}\\
    \left(1,800\;{\rm to}\;2,00,000\right)\times D(t) & {\rm for\ scenario\ B}\\
   \end{cases}
\eeq{estimateg}
The medians are 1,600 in scenario A and 15,000 in scenario B. One sees that
the medians and lower limits of the 95\% CL change by relatively small amounts,
whereas the upper limits are quite different.

\bet
\begin{center}
\begin{tabular}{|l|c||c|c|c||c|c|c|}
\hline
Geographical Cluster & D(30/3) & \multicolumn{3}{c||}{Scenario A} 
                     & \multicolumn{3}{c|}{Scenario B} \\ \cline{3-8}
        & & I(30/3) & $\cal D$(10/4) & $\cal D$(19/4) 
          & I(30/3) & $\cal D$(10/4) & $\cal D$(19/4)\\
\hline
Mumbai-Navi Mumbai-Pune & 8 & 3,700 -- 77,000 & 18 -- 61 & 39 -- 380
                            & 22,000 -- 3,00,000 & 42 -- 470 & 189 -- 18,000 \\
Indore-Ujjain           & 6 & 2,800 -- 58,000 & 13 -- 46 & 29 -- 290 
                            & 16,000 -- 2,30,000 & 32 -- 350 & 140 -- 14,000 \\
Ahmbedabad-Bhavnagar    & 5 & 2,300 -- 48,000 & 11 -- 38 & 24 -- 240 
                            & 14,000 -- 1,90,000 & 26 -- 290 & 120 -- 11,000 \\
Kolkata-Howrah          & 2 & 930 -- 19,000 & 5 -- 15 & 10 -- 96 
                            & 5,400 -- 7,50,000 & 10 -- 120 & 47 -- 4,600 \\
Karnataka North         & 2 & 930 -- 19,000 & 5 -- 15 & 10 -- 96 
                            & 5,400 -- 7,50,000 & 10 -- 120 & 47 -- 4,600 \\
\hline
\end{tabular}
\end{center}
\caption{Clusters of outbreaks, the number of reported fatalities up to
March 30, the number of infected persons inferred from this up to 30
March, and the projected number of fatalities by 10 and 19 April in
two different scenarios. All ranges quoted are 95\% confidence limits and
have been rounded to two significant digits.
Scenario A has doubling time $\tau=10$ days, and in Scenario B 
$\tau=5$ days. The projected number of
fatalities, $\cal D$, provides observational tests of the model. Note
that the distributions are very skewed and that the median lies close
to the lower limit of the very large 95\% CL.}
\eet{stats}

\bef
\begin{center}
 \includegraphics[scale=0.6]{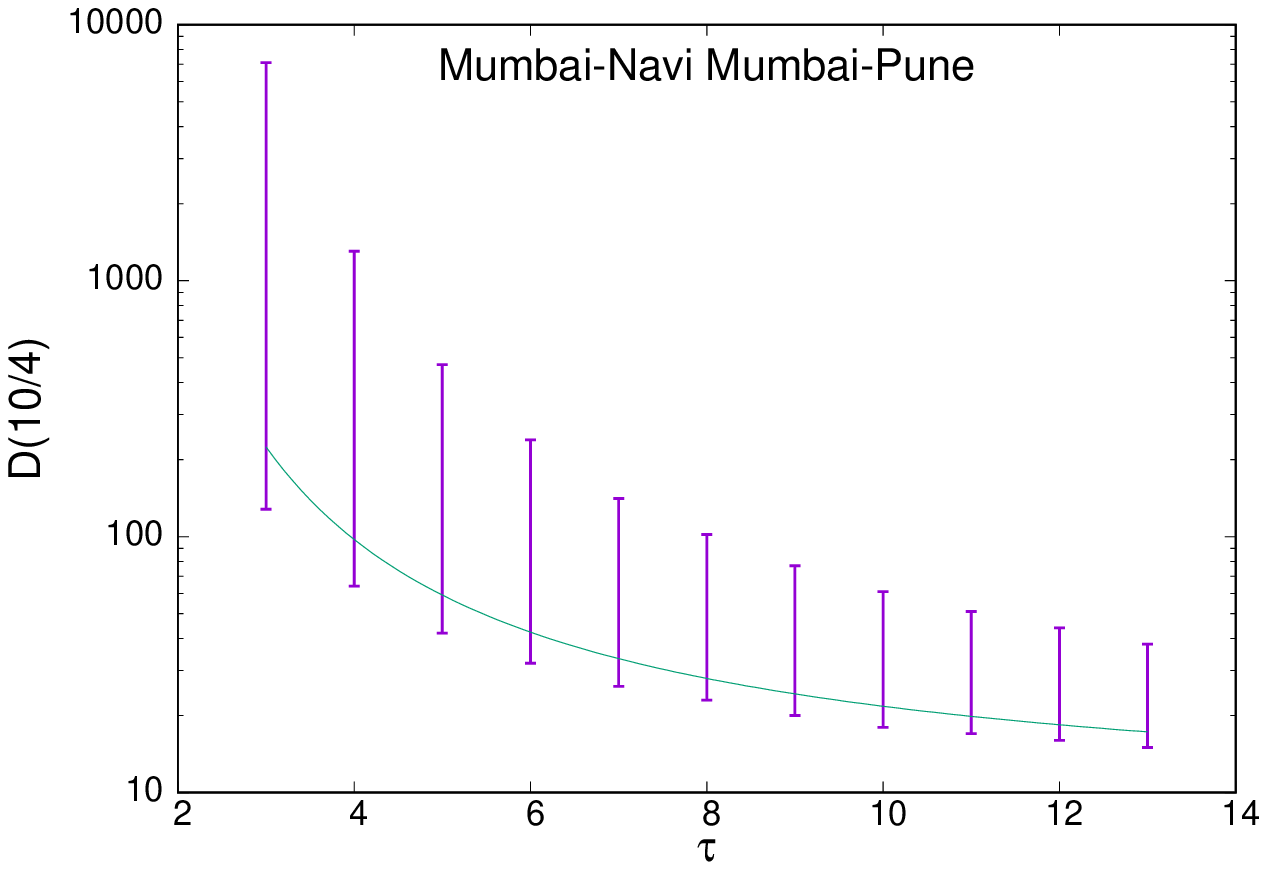}
 \includegraphics[scale=0.6]{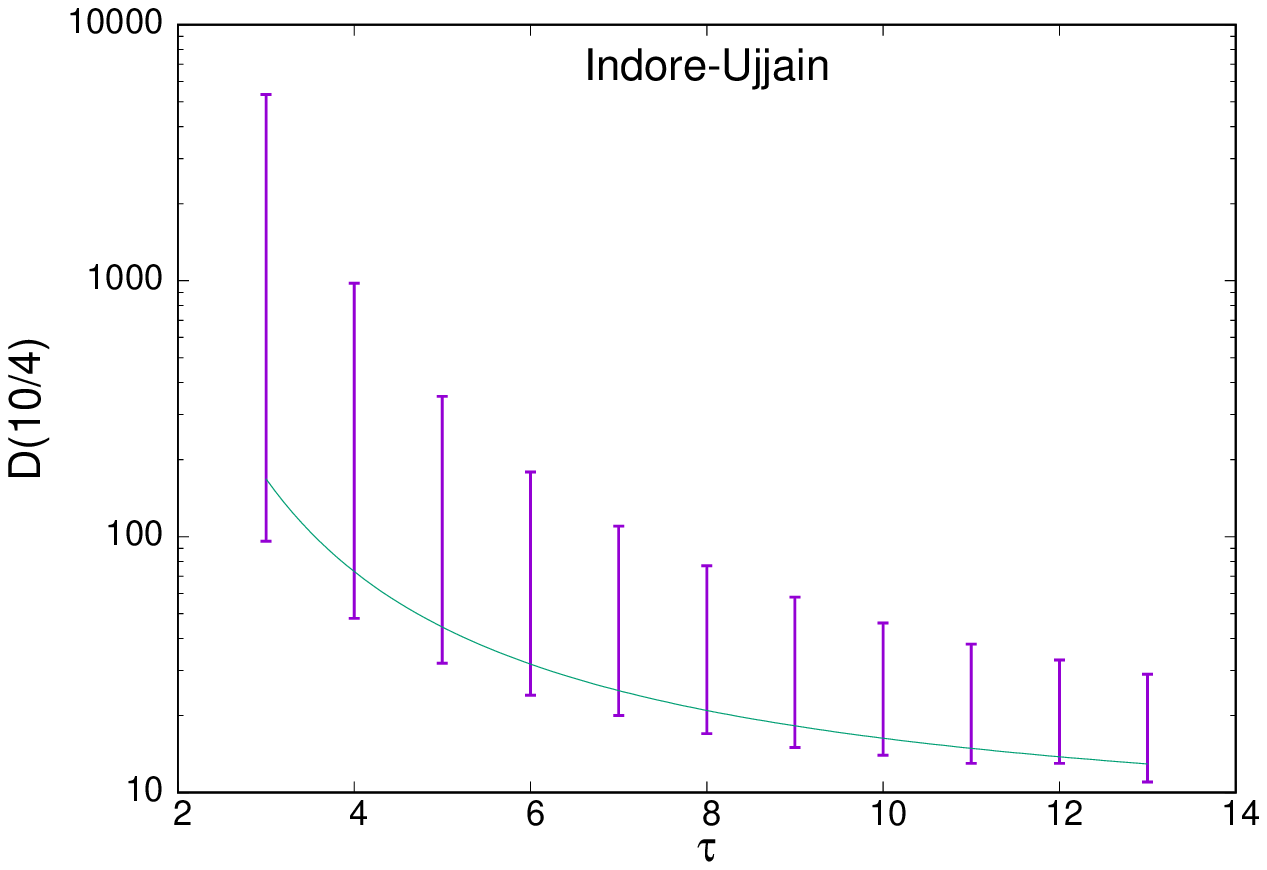}
\end{center}
\caption{The 95\% CL prediction for fatality counts ${\cal D}=I\times\ifr$
 on 10 April for different values of $\tau$ are shown as the vertical lines.
 The continuous line joins the medians of these distributions. The plot
 on the left is for the Mumbai-Navi Mumbai-Pune hot-spot and that on the
 right for the Indore-Ujjain hot-spot.}
\eef{taufind}

We list the COVID-19 hot-spots in \tbn{stats} along with estimates of $I$
and $\cal D$ on various dates for each in the two different scenarios
defined earlier. We emphasize that different geographical locations may
have different doubling times, so both scenarios may be relevant. These
are our major results. For later use we also show in \fgn{taufind} the
95\% CL predictions for $\cal D$ on 10 April in two different hot-spots in
Scenario B.  Any limits that we can extract on disease and epidemiological
parameters would help us to plan ahead for the kind of demands that may
be put on medical facilities in the near future.

\section{Discussion}

From the data on the geographical distribution of fatalities in India,
we identified four possible hot-spots for COVID-19. The clustering of
fatalities into hot-spots is an indication that there is perhaps no
general spread of the epidemic through the country, and gives hope for
partial removal of lock-down if the situation does not change.

We used the simple statistical estimator given in \eqn{estimator} to
make a prediction for the total number of infections in each of these
hot-spots on 30 March, in two scenarios. Estimating $I(t)$ is important,
especially since there is a good chance that the part of this population
which is pre-symptomatic, non-symptomatic, or has sub-clinical symptoms
are all able to communicate the disease to others.

Our predictions are exhibited in \tbn{stats}. Note that the 95\% CL
spans an enormous range, due to the spread in the input parameters,
mainly the time interval between infection and death.  Note that the
distribution of $I(t)$ is very skew, and the median is close to the
lower end of the 95\% CL. In Scenario B, the upper end of the 95\% CL
is more than 1\% of the population of Mumbai. We consider this highly
unlikely, although statistically possible. We have outlined a procedure
refine these estimates by incorporating daily data progressively into
the computation.

Given the very large values of $I(t)$ which the model predicts, medical
professionals may legitimately ask whether one sees so many respiratory
cases arriving in hospitals. Note however, that a very large fraction are
likely to be either asymptomatic, or exhibit sub clinical symptoms. In
fact \cite{ifr} simultaneously reports $\ifr$ and the fraction of infected
individuals who are hospitalized, $H$. This ratio, averaged over the
population is reported to be in the range $H=2$--4\%. In Scenario B,
therefore one may expect 440--12,000 people to arrive in a hospital.
Even among these, some may not be able to consult a physician during a
lock-down. Nevertheless, current experience strongly disfavours the upper
end of this 95\% CL range. One recalls again, that the median is close
to the lower end.  This estimate again emphasizes how important it is
to narrow the range of prediction from the model.

We are able to perform another simple estimate from these numbers. The
case fatality ratio, $\cfr$, is defined as the number of fatalities
divided by the number of cases tested. Assuming that the tests are largely
done on the people who arrive at a hospital, one can see that 
\beq
   \cfr\approx\frac{D(t)}{H I(t)} = \frac{\ifr}H.
\eeq{cfrifr}
With $\ifr=0.657$\% and $H=2$--4\%, we find $\cfr=0.16$--0.33, \ie, in
the range of 16\% to 33\%. This is precisely in the range that is seen in
the current data for India. This also means that the current policy for
testing is likely to be catching most cases which need to be hospitalized.
One hopes that with the decision to administer the faster serum test, it
might become possible to sample the larger population more effectively.

Finally we point out that there is a strong age structure to all the model
parameters, which we have ignored. This we plan to do in future. Along with
the planned Bayesian narrowing of the parameter space of COVID-19 pathology
and epidemiology, this would provide valuable inputs for more detailed
models which can be used to inform future policy.

\section{Acknowledgements}

The authors would like to thank Sandhya Koushika, Gautam Menon, and
Rahul Siddharthan for crucial inputs, and many members of the ISRC
(Indian Scientists' Response to COVID-19) mailing list for discussions.

\end{document}